\documentstyle[12pt]{article}
\newcounter{figno}                
\newcounter{label1}               
\newcounter{label2}               
\newcommand{\ltap}{\mbox{$^{_{\textstyle <}}\!\!\!\!\!_{_{\textstyle \sim}}$}}
\begin{document}
\large
\noindent
{\bf Direct experimental evidence of non-equilibrium energy sharing in dissipative collisions}

\normalsize
\vspace{.5cm}

\noindent
{\bf G. Casini, P. R. Maurenzig, A. Olmi, M. Bini, S. Calamai, F. Meucci, G. Pasquali, G. Poggi and A. A. Stefanini} 
\vspace{.2cm}

\noindent
Istituto Nazionale di Fisica Nucleare and Universit\`a di Firenze, I--50125  Florence, Italy
\vspace*{.2cm}

\noindent
{\bf A. Gobbi and K. D. Hildenbrand}
\vspace{.2cm}

\noindent
Gesellschaft f\"ur Schwerionenforschung, D--64291 Darmstadt, Germany
\vspace*{.7cm}

\begin{list} {{\bf Abstract.}} {\setlength{\leftmargin}{.4in} 
\setlength{\rightmargin}{0in} \setlength{\topsep}{0in}}

\item
Primary and secondary masses of heavy reaction products have been
deduced from kinematics and E-ToF measurements,
respectively, for the direct and reverse collisions of $^{100}$Mo 
with $^{120}$Sn at 14.1$\,A\;$MeV.
Direct experimental evidence of the correlation of energy-sharing with
net mass transfer and model-independent results on the evolution of
the average excitation from equal-energy to equal-temperature
partition are presented.
\end{list}

\vspace*{.3cm}


\vspace*{10mm}

\setlength{\baselineskip}{6.0mm}
\setlength{\topsep}{0mm}

The determination of the microscopic mechanism of energy dissipation
and energy partition between the reaction partners of a dissipative
collision has been a controversial subject of debate in the past years
[1--12]              
(for a review see \cite{TokeRev:92}).
The excitation energy sharing presents an evolution with the
inelasticity of the reaction: in quasi-elastic events the two reaction
partners reseparate with almost equal excitation energies, but with
increasing dissipation there is a trend towards equilibrium partition
({\it i.e.\ }excitation energy shared in proportion to the mass of the
fragments). 
In most cases, however, such a condition seems not to be reached even
for the largest dissipated energies \cite{Sobot:86,Peti:89,Pade:91,Lleres:93}. 
These experimental findings
can be explained by models \cite{Rand:82} which describe
the evolution of many macroscopic observables by means of stochastic
exchanges of single nucleons between the interacting nuclei.  
More refined experiments \cite{Benton:88,Wilcz:89,TokePRC:91} claimed
that the excitation energy division is correlated with the net mass
transfer, with an excess of excitation being deposited in the fragment
which gains nucleons.
Moreover, the strength of this experimental correlation seems to be
largely independent of the degree of inelasticity \cite{TokePRC:91}
and this latter result seems difficult to understand within a
stochastic nucleon exchange picture.
The present letter aims to obtain, in a model independent way, direct
experimental information on this subject.

In a previous paper \cite{CasiniPfis:91}, we used the sequential
fission to investigate the degree of equilibration between the
two reaction partners at the end of the interaction.
This was achieved with an asymmetric system
($^{120}$Sn + $^{100}$Mo at 19.1$\,A\;$MeV) in which
a given primary mass $A$ corresponds to different net mass transfers
for projectile- and target-like fragments (PLF and TLF).
The striking result was that the curves of fission probability
 $P_{\rm fiss}$ vs.\ fissioning mass for PLF and TLF do not coincide:  
for a given $A$, $P_{\rm fiss}$ for the TLF 
(which gained mass) was significantly larger than for the 
PLF, even at large Total Kinetic Energy Losses (TKEL).
However, sequential fission, while providing a tool of great 
sensitivity, did not allow to determine which variable (among those
relevant to fission, {\it e.g.\ }excitation energy, isospin, angular 
momentum, deformation) is mainly responsible for the observed effect.
Moreover, fission events form a somewhat biased
sample and may be not fully representative of all events of a
dissipative collision within a specific TKEL bin. 

In order to investigate whether similar non-equilibrium
effects are present also in the 2-body exit channel, we turned to a
different tool, namely the light particle evaporation, which depends
mainly on the excitation energy.
This procedure was applied in the past to measurement of PLF from
rather asymmetric systems studied in direct kinematics only
\cite{Benton:88,Kwiat:90,TokePRC:91} and required a detailed and not
trivial comparison between the experimental results and evaporation
calculations. 
To avoid relying on model calculations (which become increasingly
uncertain with increasing excitation energy), we aimed at comparing
not the data with a model, but directly two sets of experimental 
data. 

With an asymmetric colliding system, one might compare the two 
event samples in which reaction products of a given mass $A$ are
PLF or TLF, this fact implying different "histories"
(gained or lost nucleons).
To overcome the severe experimental difficulties (like threshold
effects, poor resolution, and critical dead layer corrections) which
impede the measurement of the TLF with sufficient accuracy, we devised
the alternative approach of measuring the secondary mass of the PLF
only, however studying the same asymmetric collision both in direct
and reverse kinematics.  This approach gives also the additional bonus
that the efficiencies for the detection of the PLF, being quite
similar for the two kinematics, practically do not affect the result
of the comparison.

This letter presents for the first time a direct experimental evidence
(based not on comparison with evaporation models, but on experimental
results only) indicating that the number of emitted nucleons depends
on the net mass transfer experienced by the primary reaction products.
This observation is strongly suggestive of a non equilibrated sharing
of the excitation energy between the two reaction partners.

Beams from the Unilac accelerator of GSI-Darmstadt were used
to study the asymmetric collision $^{100}$Mo + $^{120}$Sn at 14.1$\,A\;$MeV,
both in direct and reverse kinematics.
The moderate asymmetry of the entrance channel was chosen in order to
make sure that a common range of masses for PLF and TLF was available
even at not too large TKEL.
The experiment was based on the measurement of both the primary 
(via the kinematic coincidence method, KCM) and secondary mass
(via additional measurement of the secondary energy) of the PLF. 

Heavy ($A\geq20$) products were detected in an array of
12 position-sensitive parallel-plate avalanche detectors (PPAD),
covering about 75\% 
of the forward hemisphere \cite{CharityMo1:91,StefMo2:95}.
The FWHM resolutions of time-of-flight and position were 700 ps and
3.5 mm, respectively.
From the measured velocity vectors, primary (pre-evaporative)
quantities were deduced event-by-event with an improved version of the
KCM \cite{CasiniNim:89}.
The background of incompletely measured events of higher multiplicity
was estimated \cite{CasiniNim:89} and subtracted.

An array of 40 Si-detectors of various sizes (from 1x1 cm$^2$ at small
polar angles, up to 5x5 cm$^2$), of $300 \mu$m thickness,
was mounted behind two of the forward PPAD, so as to cover a sizeable
part of the region below and around the grazing angle, where partly
damped events are concentrated ($\theta_{\rm graz}^{\rm lab} \approx 10^{\circ}$ 
for the present collisions).
Secondary masses $A_{\rm sec}$ of the PLF were obtained
event-by-event from the energy deposited in the Si-detectors and
the time-of-flight measured by the corresponding PPAD, using an
iterative procedure which takes into account the pulse height defect
in the semiconductors and the energy loss in the PPAD and in the
dead layers. 

\addtocounter{figno}{1}
\setcounter{label1}{\value{figno}}

For various windows of TKEL (corrected for the $Q_{\rm gg}$ between
entrance and exit channel \cite{StefMo2:95}), the experimental data
were sampled in bins of reconstructed primary mass $A$ of the PLF and
the centroids of the corresponding distributions of evaporated 
mass $\Delta A = A - A_{\rm sec}$ were determined. 
The full squares and full circles in Fig.~\arabic{label1} show 
 $\Delta A$ as a function of the primary mass of the PLF in 
the direct and reverse reaction, respectively.
The presented data refer to the region TKEL $\ltap$ 500 MeV,
corresponding to partly dumped events, where PLF can be safely 
distinguished from TLF due to the strongly anisotropic angular
distributions \cite{CharityMo1:91}. 

Quantities like the primary mass $A$ and TKEL
are correlated to a certain extent with each other (being obtained
from the same velocity vectors), as well as with the secondary 
mass $A_{\rm sec}$ (via its time-of-flight). 
Moreover, the overall finite resolution (arising both from the
smearing of the particle evaporation process and from the detection
procedure) can cause systematic distortions in determining the value
of nonuniformly distributed variables (see e.g.\ the comment about the
angular distribution in \cite{CasiniNim:89} and the correction of the
mass distribution in \cite{TokeNim:90}).
As analytic corrections may be worked out only in very simple cases,
the experimental results were corrected via
extensive Monte Carlo simulations, modeling the dissipative collision
followed by an evaporative emission
in agreement with the statistical code GEMINI \cite{CharityGEM} and
incorporating as realistically as possible the response of the setup,
finite resolution effects and all known distortions of the analysis
method. The open symbols in Fig.~\arabic{label1} show the experimental
data after correction.

The most striking result resides in the different values of $\Delta A$
obtained, for a given $A$, in the direct and reverse collisions.
It has to be noted that the differences (4--6 amu)
between the evaporated masses in the two cases are much
larger than the applied corrections (1--2 amu at most).
We want also to stress that the corrections to be applied to the
experimental $\Delta A$ (and hence the corrected data points in 
Fig.~\arabic{label1}), are within errors largely independent of 
physical hypothesis ({\it e.g.\ }on energy partition), as it was
checked by repeating the Monte Carlo simulations with different
physical models. 

From the comparison between the two sets of corrected experimental
points of Fig.~\arabic{label1} one obtains information on the mechanism
of excitation energy sharing. 
The striking difference in $\Delta A$ between the two kinematic cases
can be viewed as a dependence of the excitation energy sharing on the
net mass transfer.
This behavior is put here into evidence without
recourse to statistical model calculations (our use of Monte Carlo
simulated data is limited to the correction for experimental
systematic effects).
We recall that none of the usual ways of modeling the
excitation energy sharing --- neither the equal-energy, nor the
equal-temperature scenarios, nor any combination of the two --- 
foresees the observed splitting of the correlation $\Delta A$
vs.\ $A$ into two well separated branches.

Just to clarify this point, let us focus the attention on the
symmetric exit channel, in which the two primary fragments have the
same mass number $A$=110. 
If the dinuclear system at reseparation had lost memory of its
``history'', the two excited reaction products should have de-excited
by emission of the same average number of nucleons, irrespective of
the size of the fluctuations in the internal degrees of freedom.
Thus the observed difference in $\Delta A$ indicates a sizeable
deviation from equilibrium at the end of the interaction phase. 
Actually, due to the enhanced sensitivity of the particle evaporation
process to the excitation energy of the emitter (with respect to other 
internal variables like isospin, angular momentum \cite{Sobot:86}  
or deformation), this experimental result is a proof of a sizeable
deviation from equilibrium in the excitation energy sharing. 
Neglecting pre-equilibrium emission and evaporation from the dinucleus
during the interaction phase (which are small at these bombarding
energies \cite{Wile:89,CharityMo1:91}), one can estimate the
mean excitation energy $\epsilon$ removed per evaporated nucleon.
Dividing the central value of the TKEL bin by the sum of the masses
evaporated by the PLF with A=110 in the direct and reverse reaction,
one obtains, at all TKEL, values of $\epsilon$ ($\approx$ 11--12 MeV)
which are in good agreement with the 12--13 MeV predicted by GEMINI.
Thus one roughly estimates that, of two nuclei of primary mass A=110,
the one obtained by a gain of 10 nucleons (PLF in the direct reaction)
should be about 50--60 MeV more excited than the one obtained by
removal of 10 nucleons (PLF in the reverse reaction). 

\addtocounter{figno}{1}
\setcounter{label2}{\value{figno}}

More quantitatively, one can build the ratio
\begin{equation}
R =(\Delta A^l_{110}-\Delta A^h_{110})/(\Delta A^l_{110}+\Delta A^h_{110})
                                                \label{R}
\end{equation}
where $\Delta A^l_{110}$ ($\Delta A^h_{110}$) is the total evaporated
mass for nuclei with primary mass $A$=110, originating from the
entrance channel light (heavy) nucleus and measured as PLF
in direct (reverse) kinematics. Fig.~\arabic{label2}(a) shows
the so defined $R$ as a function of TKEL. 
 $R$ is an estimate of the excitation-energy asymmetry
 $(E^{\ast l}_{110}-E^{\ast h}_{110})/ (E^{\ast l}_{110}+E^{\ast h}_{110})$,
being $E^{\ast l}_{110} = \epsilon \Delta A^l_{110}$ 
($E^{\ast h}_{110} = \epsilon \Delta A^h_{110}$) the excitation energy
of nuclei with $A$=110, originating from the entrance channel
light (heavy) nucleus.

In the present experiment, also the average partition of excitation
energy between the two reaction partners can be deduced --- in a
substantially model-independent way --- from the number of nucleons
emitted in case of no net mass transfer.
Fig.~\arabic{label2}(b) shows the ratio
\begin{equation}
 C =(\Delta A^h_{120}-\Delta A^l_{100})/( \Delta A^h_{120}+\Delta A^l_{100})
                                                              \label{C}
\end{equation}
as a function of TKEL, where $\Delta A^l_{100}$ ($\Delta A^h_{120}$)
is the total evaporated mass for nuclei of primary mass A=100 and 120
originating from the entrance channel light and heavy nucleus,
respectively.
 $C$ is an estimate of the excitation energy partition
 $(E^{\ast h}_{120}-E^{\ast l}_{100})/(E^{\ast h}_{120}+E^{\ast l}_{100})$,
when assuming a common value $\epsilon$ for the average energy
necessary to evaporate a single nucleon from the two reaction partners
(we verified with GEMINI that indeed the actual values of $\epsilon$
for A=100, 120 differ by no more than 1$\%$ to 3$\%$  
when $E^{\ast}$ ranges from 50 to 250~MeV).

Our data show that for the exit channel without net mass transfer the
total excitation energy is initially almost equally shared between the
fragments ($C \approx$ 0 at small TKEL) and that the approach to the
equilibrium partition (mass ratio, shown by the dotted line in 
Fig.~\arabic{label2}(b)) is very slow with increasing TKEL.
In spite of the limited range of $C$ values spanned between these two 
extremes (less then 10$\%$, 
due to the moderate mass asymmetry of the collision), we find that
even at the highest explored TKEL a partition consistent with full
thermal equilibrium is not reached.
This average behavior is in agreement with other experimental results
and compatible, by itself, with model descriptions based on
the exchange of independent nucleons during the contact phase. 

A linear dependence of the energy partition on the net mass
transfer had been proposed by Toke {\it et al.\ }\cite{TokePRC:91}:
\begin{equation}
 E^{\ast}_{\rm PLF}(A)/E^{\ast}_{\rm tot}=C_{T}+R_{T}\,(A-A_{\rm beam}) 
                         \label{EToke}
\end{equation}
where the excitation energy of a PLF of mass $A$ was given in terms of
the TKEL-dependent parameters $C_{T}$ and $R_{T}$.

Using the parameters defined in Eq.\ (\ref{R}) and (\ref{C}), we can
write Eq.\ (\ref{EToke}), for the products deriving from the 
original light or heavy colliding nucleus:
\begin{equation}
 \frac{E^{\ast \; l,h}(A)}{E^{\ast}_{\rm tot}} = \frac{1}{2} + 
     \frac{C}{A_{\rm dif}} \left(A - \frac{A_{\rm tot}}{2}\right)
    +   \frac{R}{A_{\rm dif}} \left(A - A^{l,h}_0\right)       
                         \label{Enoi}
\end{equation}
where $A^l_0$ ($A^h_0$) is the lighter (heavier) mass
between $A_{\rm beam}$ and $A_{\rm target}$ in the entrance channel, 
 $A_{\rm tot}=A^l_0+A^h_0$ and  $A_{\rm dif}=A^h_0-A^l_0$.
Our notation has the advantage of making evident that there is a 
mass dependent term (containing $C$) which simply describes the
dependence of excitation energy on mass ({\it e.g.}, in case of
thermal equilibrium, $C = A_{\rm dif}/A_{\rm tot}$ leading to a trivial
proportionality to the mass of the nucleus).  
However, only the term containing $R$ truly represents a
net-mass-transfer dependent term and it is responsible for the
splitting of the correlation into two distinct branches.  
The slope parameter $R_T$ of Eq.\ (\ref{EToke}) mixes the two terms as
it comes out to be  $R_T \equiv (C+R)/A_{\rm dif}$. 
We want to stress that the experimental decomposition of $R_T$ in the
two contributions was possible only in the present experiment, due to
the measurement of the PLF in the direct and reverse kinematics.
Previous experiments, measuring the evaporated mass of PLF in one
kinematic case only, could attempt such a decomposition  
only in a model-dependent way.

The observed correlation between ``evaporated mass'' and net mass
transfer, which is here evidenced without need of model calculations,
strongly suggests that there is no complete equilibrium between the
two reaction partners.
The persisting strength of such a correlation even at high TKEL 
is a strong challenge for a microscopic description. 
Within a nucleon exchange picture, some correlation might arise 
from a possible donor-acceptor intrinsic asymmetry in the excitation
energy deposition caused by the exchange of a single nucleon. 
However, with increasing dissipation, the larger and larger number of
independent nucleon exchanges should almost wash out the correlation.
Toke {\it et al.\ }\cite{TokePRC:91} emphasized the surprising
constancy of the donor-acceptor intrinsic asymmetry $\eta$ which they
deduced from a re-analysis of the data of Ref.\ \cite{Benton:88}.
Following Ref.\ \cite{TokePRC:91}, which employed, as usual, the
experimental mass variances $\sigma^2_A$ as an estimate of the total
number of exchanges, we obtain unreasonable values from our data
($\eta$ comes out to be greater than unity and increases with
increasing TKEL), also because of the very rapid increase
of $\sigma^2_A$ with TKEL.  
In the spirit of Ref.\ \cite{TokePRC:91}, also this fact points to a
failure in the present description of the nucleon transfer
process at larger bombarding energies. 
Indeed, with increasing TKEL and bombarding energy, other effects might
come into play, which cannot be described simply with the elementary
process of single nucleon transfer across a window.  
For example, remaining in the framework of one-body dissipation picture,
the rapid increase of $\sigma^2_A$ with TKEL could be reconciled with
a smaller number of exchanges if a relevant contribution comes from
the transfer of clusters of nucleons.
Alternatively, a relevant role could be played by collective effects,
such as formation and rupture of a neck during the collision. 
Both of these suggestions require complete and precise theoretical
calculations. 

In conclusion, the existence of a mechanism which correlates the
evaporated mass --- and hence the excitation energy sharing --- with the
net mass transfer has been evidenced in a model independent way.
This experimental finding seems difficult to reconcile with existing
models based on stochastic exchanges of singles nucleons and 
calls for a better theoretical understanding of the microscopic
interaction mechanism of heavy nuclei.

We wish to thank the staff of the Unilac accelerator for their
skillfulness in delivering high quality Mo and Sn beams pulsed 
with good time structure, as well as
P. Del Carmine and F. Maletta for their valuable
support in the preparation of the experimental set-up.

\vspace*{15mm}

\large
{\bf Figure Captions}
\normalsize

\vspace*{3mm}

\begin{list} {} {\setlength{\rightmargin}{\leftmargin} }

\parbox{150mm}{\setlength{\baselineskip}{6mm}
\item[Fig.~\arabic{label1}]
The full squares (circles) show the experimental average number of
evaporated nucleons $\Delta A$ as a function of the primary mass $A$ 
of the PLF in the direct (reverse) reaction, for windows of TKEL. 
 $\Delta A$ is the difference between the primary mass and the
centroid of the corresponding distribution of secondary masses.  
The open symbols show the the experimental
data after correction for the response of the setup, finite
resolution effects and distortions of the analysis.
}
\vspace{4mm}

\parbox{150mm}{\setlength{\baselineskip}{6mm}
\item[Fig.~\arabic{label2}]
(a): Asymmetry in the total evaporated mass
 $R =(\Delta A^l_{110}-\Delta A^h_{110})/ (\Delta A^l_{110}+\Delta A^h_{110})$ 
for nuclei of primary mass $A$=110, originating from
the entrance channel light and heavy nuclei, as a function of TKEL. 
 $R$ is an estimate of the excitation-energy asymmetry
due to the net gain or loss of nucleons.\\
(b): Asymmetry in the evaporated mass
 $C =(\Delta A^h_{120}-\Delta A^l_{100})/(\Delta A^h_{120}+\Delta A^l_{100})$
for nuclei with A=100 and 120 originating from the entrance channel
light and heavy nucleus, respectively, as a function of TKEL.
 $C$ is an estimate of the excitation energy partition in absence of
net mass transfer:
 $C=0$ indicates equipartition of excitation (equal energy
sharing), the dotted line shows the value expected for thermal
equilibrium (equal temperature sharing). 
}
\end{list}

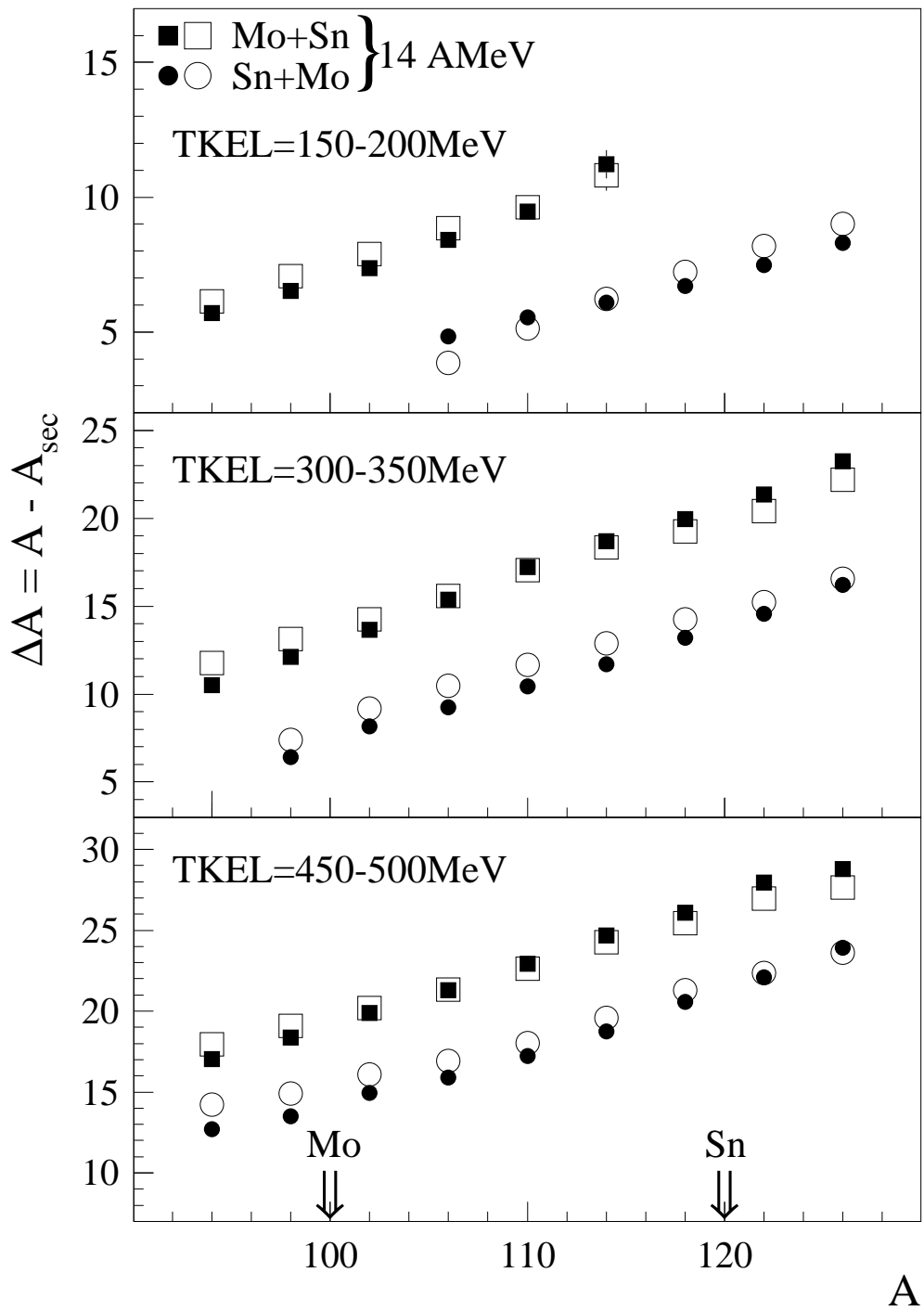
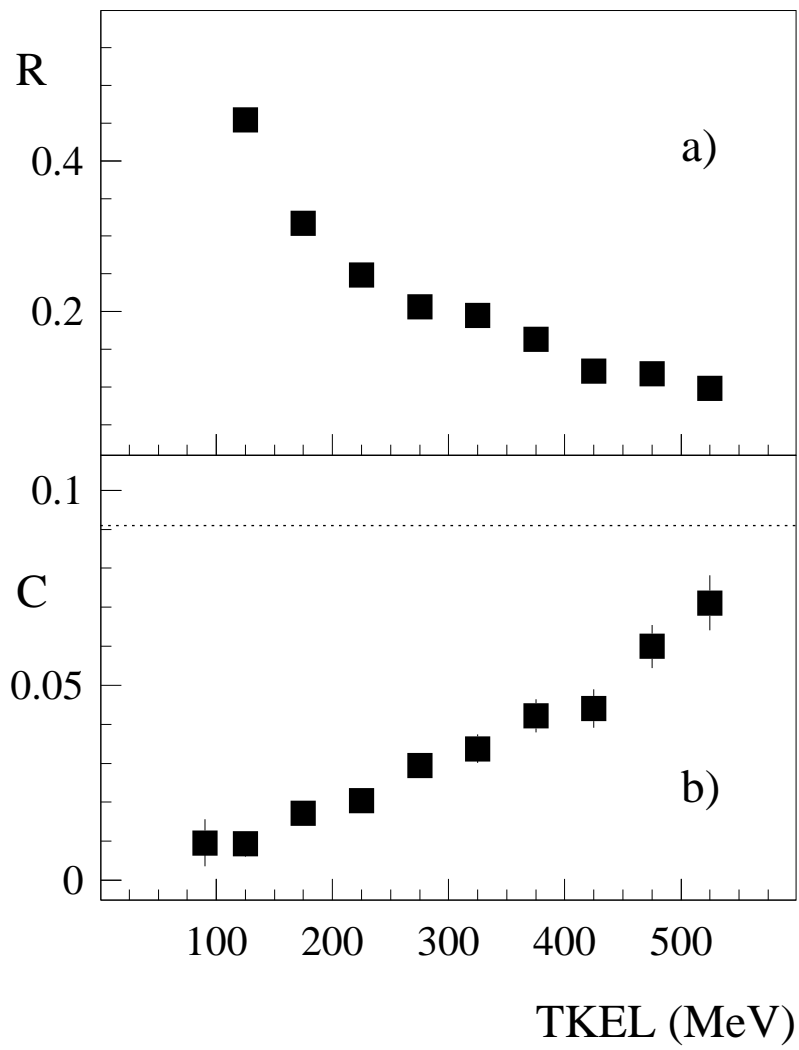
\end{document}